\documentclass[pra,twocolumn,nofootinbib,superscriptaddress]{revtex4-2}
\usepackage{xspace,amsmath,amsfonts,amsthm,amssymb,amsbsy}
\usepackage{graphicx}
\usepackage{amssymb}
\usepackage{color}
\usepackage{psfrag}
\usepackage{ifsym}
\usepackage{epstopdf}
\usepackage[usenames,dvipsnames]{xcolor}
\usepackage[colorlinks=true,citecolor=Cerulean,linkcolor=RubineRed,urlcolor=Cerulean]{hyperref}
\usepackage{bbold}
\usepackage{cancel}
\usepackage{braket}
\usepackage{rotating} 
\usepackage{mathrsfs}
\definecolor{orchid}{rgb}{0.85, 0.44, 0.84}
\definecolor{pink}{rgb}{1,0.078,0.57}

\definecolor{green}{rgb}{0,0.7,0.9}
\newcommand{\aurelia}{\color{black}}

\newcommand{\pablo}{\color{black}}

\newcommand{\pma}[1]{{\textcolor{black}{#1}}}

\newcommand{\pb}[1]{\{#1\}_\textsc{p}}

\newcommand{\dg}{^{\dagger}}
\newcommand{\Tr}{\mathrm{Tr}}

\def\dg{^\dagger}
\newcommand{\beq}{\begin{equation}}
	\newcommand{\eeq}{\end{equation}}
\newcommand{\beqa}{\begin{eqnarray}}
	\newcommand{\eeqa}{\end{eqnarray}}

\newcommand{\avr}[1]{\langle #1 \rangle_\rho}
\newcommand{\av}[1]{\langle #1 \rangle}

\newcommand{\avs}[1]{\langle #1 \rangle} 

\newcommand{\be}[1]{\mathbf{#1}}
\newcommand{\mc}[1]{\mathcal{#1}}
\newcommand{\mr}[1]{\mathrm{#1}}
\newcommand{\ts}[1]{\textsc{#1}}

\newcommand{\msc}[1]{\mathscr{#1}}
\newcommand{\ha}[1]{\hat{#1}}

\newcommand{\mbb}[1]{\mathbb{#1}}

\begin{document}

	\title{Stochastic Operator Variance: an observable to diagnose noise and scrambling
	}
	
		\author{Pablo Martinez-Azcona}
		\email{pablo.martinez@uni.lu}
	\affiliation{Department of Physics and Materials Science, University of Luxembourg, L-1511 Luxembourg}
	
	\author{Aritra Kundu}
	\affiliation{Department of Physics and Materials Science, University of Luxembourg, L-1511 Luxembourg}
	
	\author{Adolfo del Campo}
	\affiliation{Department of Physics and Materials Science, University of Luxembourg, L-1511 Luxembourg}
	\affiliation{Donostia International Physics Center, E-20018 San Sebasti\'an, Spain}
	
	\author{Aur\'elia Chenu}
	\email{aurelia.chenu@uni.lu}
	\affiliation{Department of Physics and Materials Science, University of Luxembourg, L-1511 Luxembourg}

	
	\begin{abstract}
		
		\pma{Noise is ubiquitous in nature, so it is essential to characterize its effects. Considering a fluctuating Hamiltonian, we introduce an observable, the stochastic operator variance (SOV), which measures the spread of different stochastic trajectories in the space of operators. The SOV obeys an uncertainty relation and allows finding the initial state that minimizes the spread of these trajectories. We show that the dynamics of the SOV is intimately linked to that of out-of-time-order correlators (OTOCs), which define the quantum Lyapunov exponent $\lambda$.} Our findings are illustrated analytically and numerically in a stochastic Lipkin-Meshkov-Glick (sLMG) Hamiltonian undergoing energy dephasing.
		
	\end{abstract}
	
	\maketitle
	
	\pma{Any realistic quantum system inevitably experiences some fluctuations induced by its surrounding environment.} 
	Current quantum technologies are limited by the action of this noise, motivating the pragmatic focus on Noisy-Intermediate-Scale Quantum (NISQ) devices \cite{Preskill18nisq,Bharti22}.
	In any experimental setting, tunable parameters such as Hamiltonian coupling constants may exhibit fluctuations due to interactions with the environment \cite{budini_quantum_2001, chenu_quantum_2017, gardiner_quantum_2004}. 
	In this context, the dynamics of an ensemble of noisy realizations can be described in terms of the noise-averaged density matrix, which evolves according to a master equation  describing non-unitary evolution \cite{budini_quantum_2001,chenu_quantum_2017,xu_extreme_2019, kiely_exact_2021}. Alternatively, noise can be utilized as a resource for the quantum simulation of open systems \cite{chenu_quantum_2017}. The study of fluctuations in noisy quantum systems is also connected to free probability \cite{bernard_open_2019, bernard_dynamics_2022, hruza_coherent_2022}.
	
	The quest for understanding noise in chaotic systems has recently led to a flurry of activities exploring the signatures of quantum chaos when the dynamics is no longer unitary \cite{haake_quantum_2010,Liu2017,xu_extreme_2019,Can19,del_campo_decoherence_2020,sa_complex_2020,SaProsen20,xu_thermofield_2021,GarciaGarcia22}.
	Out-of-time-order correlators (OTOCs) offer an important diagnostic tool, which was  initially proposed in the theory of superconductivity \cite{larkin_quasiclassical_1969}. Their use experienced renewed interest in defining a quantum analog of the Lyapunov exponent \cite{kitaev_hidden_2014, richter_semiclassical_2022}, which measures the exponential sensitivity to the initial conditions in chaotic systems and is universally bounded by the system's temperature \cite{maldacena_bound_2016}. The existence of a positive Lyapunov exponent classically is a necessary but not sufficient condition for the system to be chaotic---e.g. \cite{strogatz_nonlinear_2000, wimberger_nonlinear_2014}. Similarly, the exponential growth of the OTOC is not a sufficient signature for quantum chaos but rather indicates \textit{scrambling} \cite{xu_does_2020, rozenbaum_early-time_2020}. \pma{OTOCs have been studied experimentally
		\cite{li_measuring_2017, garttner_measuring_2017, joshi_quantum_2020, green_experimental_2022,nie_detecting_2019, vermersch_probing_2019}} and in open systems where their evolution is changed by dissipation \cite{syzranov_interaction-induced_2019, zanardi_information_2021, swingle_resilience_2018, richter_semiclassical_2022, zhang_information_2019}. 
	\begin{figure}
		\centering
		\includegraphics[width = .6\linewidth]{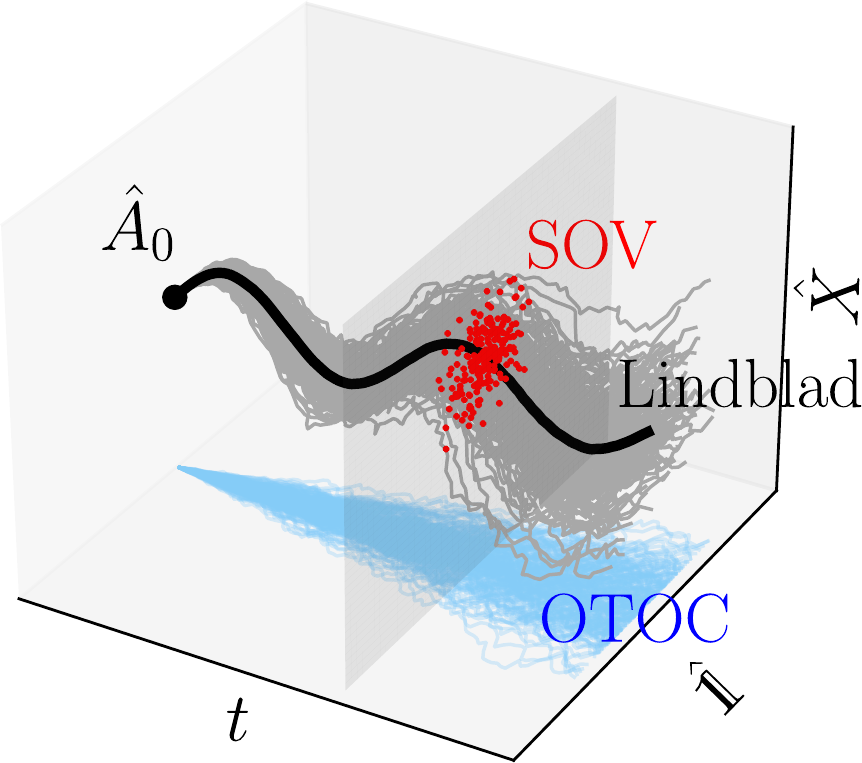}
		\caption{\textbf{The SOV-OTOC connection.} \pma{Illustration of the stochastic operator variance (SOV) and its connection to} the out-of-time-order correlator (OTOC). An operator $\ha A$ evolves through different realizations (gray) of a stochastic Hamiltonian, as illustrated by its projections over the identity $ 
			\ha{\mathbb 1}$ and another operator $\ha X$. The noise-averaged evolution (black) follows Lindblad dissipative dynamics. The SOV $\Delta \ha A_t^2$ characterizes the deviation of different trajectories (red). Its projection over the identity (blue) \pma{corresponds to} the evolution of the OTOC. }
		\label{fig:sOTOC_vis}
	\end{figure}
	
	In this Letter, we consider the dynamics generated by a stochastic Hamiltonian and go beyond \pma{an average description of the state by introducing the Stochastic Operator Variance (SOV), an observable that characterizes the spread of stochastic trajectories of any operator.} This notion is directly relevant to experiments, particularly in NISQ devices subject to various sources of noise. \pma{We compute the evolution of the SOV and show  that it obeys a generalized uncertainty relation. It allows to identify the initial state that minimizes the deviation from Lindblad dynamics at long times.} Surprisingly, we also find 
	that \pma{the SOV evolution relates to that of an OTOC}, which connects fluctuations of the system with scrambling. This is pictorially represented in
	Fig.~\ref{fig:sOTOC_vis}. 
	We illustrate our results in a  stochastic generalization of the  Lipkin-Meshkov-Glick (sLMG) model.
	The LMG model \cite{lipkin_validity_1965} describes 
	an Ising spin chain with infinite range interactions and exhibits scrambling from an unstable fixed point \cite{pilatowsky-cameo_positive_2020, pappalardi_scrambling_2018, xu_does_2020}. It can be realized experimentally with trapped ions \cite{islam_onset_2011} and is amenable to dynamical control techniques such as shortcuts to adiabaticity \cite{campbell_shortcut_2015}. 
		\pma{Considering fluctuations in the energy scale, we compute the SOV in this model, show its connection to the OTOC, and use it to characterize the Lyapunov exponent in the classical limit.}

	\emph{The stochastic operator variance (SOV)}.---
	Let us consider a system evolving under a Hermitian Hamiltonian $\ha H_0$ and subject to classical noise $\xi_t$ modulating the coupling constant of  a Hermitian  operator $\ha L$, i.e., 
	\begin{equation}\label{eq:Hstoc}
		\ha H_t = \ha H_0 + \sqrt{2 \gamma} \, \xi_t  \ha L, 
	\end{equation}
	where $\gamma$ measures the coupling strength between the system and noise---we set $\hbar = 1$. The stochastic process is taken as real Gaussian white noise, that is,  $\avs{\xi_t}= 0$ and $\avs{\xi_t \xi_{t'}}=\delta(t-t')$. We will represent the stochastic averages by $\avs{\bullet}$; quantum expectation values will be written explicitly taking the trace over the state density matrix, $\Tr(\bullet \rho)$.
	We introduce the Wiener process $\mr d W_t \equiv \xi_t \mr dt$, which is  convenient to deal with the formal treatment of stochastic differential equations (SDEs). It obeys It\=o's rules $\mr d W_t^2 = \mr dt$ with vanishing higher-order terms, $\mr dt \,  \mr dW_t = \mr dW_t^{k+1}  =\mr dt^k=0\; \forall k >1$ \cite{gardiner_handbook_1985}.
	{\pablo  Over a time increment $\mr dt$, the evolution of a Hermitian operator $\ha A$ in the Heisenberg picture is  $\ha A_{t-\mr dt}= \ha U_{\mr dt}\dg \ha A_t \ha U_{\mr dt}$ (see App. \ref{sec1}), since operators in the Heisenberg picture evolve backward in time \cite{gammelmark_past_2013, warszawski_solving_2020}. The associated propagator  reads (see App. \ref{sec1})
		\begin{equation}
			\ha U_{\mr dt}= e^{- i \ha H_0 \mr dt - i \sqrt{2 \gamma} \mr d W_{t-\mr dt}\ha L}. 
		\end{equation}  
		Expanding the propagator according to those rules and introducing the backward differential $\mr d \ha A_t = \ha A_{t-\mr dt} - \ha A_{t}$ \cite{gammelmark_past_2013, warszawski_solving_2020} yields the SDE for the evolution of $\ha A_t$ under the  stochastic Hamiltonian \eqref{eq:Hstoc},}
	\begin{equation}\label{eomAt}
		\mr d \ha A_t = \mc L\dg[\ha A_t] \mr dt + i \sqrt{2 \gamma} [\ha L, \ha A_t] \mr d W_{t - \mr dt},
	\end{equation}
	where $\mc L\dg[\ha A_t]=i [\ha H_0, \ha A_t] - \gamma [\ha L, [\ha L, \ha A_t]]$ is the adjoint Lindbladian. 
	Upon averaging \eqref{eomAt}, all the linear terms in $\mr d W_{t- \mr dt}$ vanish \cite{gardiner_handbook_1985} and
	we find that the noise-averaged operator evolves with an adjoint Lindblad equation $\mr d_t \av{\ha A_t} = \mc L\dg[\av{\ha A_t}]$. This corresponds to the standard evolution of an observable in an open quantum system with a Hermitian jump operator $\ha L$ \cite{breuer_theory_2002}. 
	The formalism described so far has been introduced in \cite{budini_quantum_2001} and used in \cite{chenu_quantum_2017} to engineer long-range and many-body interactions. Here, we focus on the stochastic variance of an observable. 
	
	In order to find the variance, the second stochastic moment $\av{\ha A_t^2}$ is needed. \pma{Considering the 
	reference} operator to be $\ha A^2$ instead of $\ha A$, \pma{one finds that its evolution} follows the Lindblad equation, $\mr d_t \av{\ha A_t^2}=\mc L\dg[\av{\ha A_t^2}]$. 
	Recall that the average is over realizations of the noise and that $\avs{\ha A_t^2}$ is still an operator acting on the Hilbert space. 
	Subtracting $\mr d_t\avs{\ha A_t}^2{= \avs{\ha A_t} \mc L\dg[\avs{\ha A_t}]+ \mc L\dg[\avs{\ha A_t}]}\avs{\ha A_t}= \mc L\dg[\av{\ha A_t}^2]+2 \gamma [\ha L, \av{\ha A_t}]^2 $  from both sides (see App. \ref{sec1}), we  find the evolution of the SOV, defined as $\Delta \ha A^2_t = \avs{\ha A_t^2}-\avs{\ha A_t}^2$, to be given by 
	\begin{equation}\label{eqVar}
		\frac{\mr d(\Delta \ha A^2_t)}{\mr dt}= \mc L\dg[\Delta \ha A_t^2]-2 \gamma [\ha L, \av{\ha A_t}]^2 .
	\end{equation}
	The SOV $\Delta \ha A_t^2$ is an \pma{observable} that characterizes the deviation of any (stochastic) operator $\hat{A}_t$ from the noise-averaged operator in a stochastic evolution governed by the  Hamiltonian \eqref{eq:Hstoc}---see Fig.~\ref{fig:sOTOC_vis} for a scheme {\aurelia and  App. \ref{sec4} for a quantitative illustration}. Although its equation of motion depends on out-of-time-order terms like $\ha L \av{\ha A_t} \ha L \av{\ha A_t}$, it can easily be computed from the evolution of $\av{\ha A_t}$ and $\av{\ha A_t^2}$. Indeed, the SOV evolves as 
	\beq \label{defSOV}
	\Delta \ha A_t^2 \equiv \avs{\ha A_t^2}-\avs{\ha A_t}^2 = e^{\mc L\dg t} [\ha A^2] - (e^{\mc L\dg t} [\ha A])^2.
	\eeq 
	The average $\av{\ha A_t}$ follows the standard Lindblad dissipative dynamics from which the individual trajectories deviate as dictated by the variance $\Delta \ha A_t^2$. 
	{\pablo Since the SOV is Hermitian, it is an observable. It differs from the quantum variance of an operator $\ha A$  over a state $\rho_t$, commonly defined as $\mr{Var}(\ha A, \rho_t) = \Tr(\ha A^2 \rho_t) - \Tr(\ha A \rho_t)^2 $. For an initially pure state, this variance reads  $\mr{Var}(\ha A_t, \psi_0)=\braket{\psi_0|e^{\mc L\dg t}[\ha A^2]|\psi_0}-\braket{\psi_0|e^{\mc L\dg t}[\ha A]|\psi_0}^2$. Such quantum variance cannot be obtained from the expectation value of an observable since it is nonlinear in the state $\ket{\psi_0}$. The difference between the SOV \eqref{defSOV} evaluated on a pure state and the quantum variance  reads
		$\braket{\psi_0|\Delta \ha A_t^2 |\psi_0} - \mr{Var}(\ha A_t, \psi_0) = \braket{\psi_0|e^{\mc L\dg t}[\ha A]\ha{\mbb Q}e^{\mc L\dg t}[\ha A]|\psi_0}$, 
		where $\ha{\mbb Q}= \mbb 1 - \ket{\psi_0}\bra{\psi_0}$ is a projection operator on the complementary subspace of $\ket{\psi_0}\bra{\psi_0}$. Therefore, the SOV contains the information of the quantum variance plus the contribution from the projector over the complementary subspace, as the term governing the recombination of decay products in unstable systems  \cite{beau_nonexponential_2017}.
		
		A related notion of variance---without the stochastic 
		interpretation---has been introduced in the theory of positive definite matrices \cite{bhatia_positive_2009}, which provides us with tools to formally characterize the SOV. First, the SOV is positive semidefinite, $\Delta \ha A_t^2\geq 0$. This can be shown from Kadison's inequality \cite{kadison_generalized_1952} which ensures that $e^{\mc L\dg t}[\ha A^2]\geq (e^{\mc L\dg t}[\ha A])^2$ for a positive and unital map, $e^{\mc L\dg t}[\mbb 1]=\mbb 1$. Second, our formalism leads to a generalization of the Robertson-Schr\"odinger uncertainty principle \cite{robertson_uncertainty_1929, schrodinger_zum_1930}. Indeed, considering two operators $\ha A, \; \ha B$ and their SOVs, we  show in App. \ref{sec1} that 
		\begin{align} \label{SOV_unc_rel}
			\Tr(\Delta \ha A_t^2 \rho_0)\Tr(\Delta \ha B_t^2 \rho_0)&\geq |\Tr(\Delta \widehat{AB}_t \rho_0)|^2, \\ 
			&\geq \frac{1}{4}\left( D_+^2(\ha A,\ha B) -D_-^2(\ha A,\ha B)\right) ,\notag
		\end{align}
		where $\Delta \widehat{AB}_t = e^{\mc L\dg t}[\ha A \ha B]-e^{\mc L\dg t}[\ha A]e^{\mc L\dg t}[\ha B]$ is the stochastic operator covariance of $\ha A$ and $\ha B$. 
		The quantity $D_\eta( \ha A, \ha B)\equiv \Tr\Big( e^{\mc L\dg t}\big([\ha A, \ha B]_\eta\big)\rho_0 -[ e^{\mc L\dg t}(\hat{A}),e^{\mc L\dg t}(\hat{B}) ]_\eta) \rho_0 \Big)$ measures the difference between evolving the commutator ($\eta=-$) and anti-commutator ($\eta=+$) as a whole or each separately (see App. \ref{sec1}). Note that $D_+$ is purely real and $D_-$ is purely imaginary, so $D_+^2 - D_-^2 \geq 0$. For $\ha A=\ha B$, the uncertainty relation is saturated and Eq. \eqref{SOV_unc_rel} becomes an equality.
		
		\emph{Bi-partite interpretation}---Many quantum properties, such as entanglement or scrambling, are best understood in a bipartite system \cite{zanardi_entanglement_2001, styliaris_information_2021, anand_brotocs_2022, pappalardi_quantum_2022}.  
		The SOV can actually be interpreted analogously: consider a doubled Hilbert space $\mathscr H \otimes \mathscr H$, with an operator $\ha A$ living in each of the copies of the Hilbert space, denoted as $\mathscr H_1$ and $ \mathscr H_2$. Using  the swap operator $\mbb S$ \cite{anand_brotocs_2022, pappalardi_quantum_2022}, which introduces an interaction between the Hilbert spaces and is defined as $\mbb S (\ket{i}_1\ket{j}_2)= \ket{j}_1\ket{i}_2$,  
		the product between operators can be interpreted as an operation over a doubled Hilbert space, namely $\ha X \ha Y = \Tr_{\mathscr H_2}\big((\ha X \otimes \ha Y) \mbb S\big)$---see details in App. \ref{sec3}. Thus, the first term in the SOV \eqref{defSOV} can be interpreted as first letting the operators $\ha A$ interact to form $\ha A^2$ in the bipartite system and then letting that operator evolve under the dissipative evolution of a single bath, $e^{\mc L\dg t }[\ha A^2]$. By contrast, the second term corresponds to letting 
		each of the uncoupled systems evolve with their own bath and then making them interact 
	    at time $t$. As expected, the difference between these terms is positive since $\Delta \ha A_t^2\geq 0$, and the first protocol always suffers less decoherence. 
	}

	\emph{The SOV-OTOC connection}---Remarkably, the expectation value of \eqref{eqVar} over the completely-mixed state, $\ha \rho = \ha{\mbb 1}/N$, gives a dissipative version of the OTOC, namely 
	\beq \label{eq:otoc}
	\frac{1}{N} \frac{\mr d \Tr(\Delta \ha A_t^2)}{\mr dt} = - \frac{2 \gamma}{N}\Tr([\ha L, \av{\ha A_t}]^2). 
	\eeq
	OTOCs are typically defined from two operators as $C_t = -\Tr([\ha B_0, \ha A_t]^2)/N$, and measure the exponential sensitivity on initial conditions in quantum chaotic systems \cite{kitaev_hidden_2014}. 
	Indeed, in a quantum system with scrambling, one expects $C_t \sim \epsilon e^{\lambda_\textsc{q} t}$ in the time window  $t_s \ll t \ll t_\ts{e}$
	between the saturation time of two-point functions, $t_s\sim 1/\lambda_\textsc{q} $, and that of the OTOC, known as the Ehrenfest time, $t_\textsc{e}\sim \ln(\hbar^{-1})/\lambda_\textsc{q} $ \cite{maldacena_bound_2016}. 
	The main difference in our setting is that the evolved operator follows dissipative dynamics, $e^{\mc L\dg t} [\ha A]$, instead of unitary evolution, $e^{i \ha H t} \ha A e^{- i \ha H t}$.  
	The connection between this OTOC and the SOV is pictorially shown in Fig.~\ref{fig:sOTOC_vis}. It can be used to compute the Lyapunov exponent through 
	\begin{equation}\label{eqExplambdat}
		C_t = \frac{1}{2 \gamma N}\frac{\mr d \Tr(\Delta \ha A_t^2)}{\mr d t}\sim  \epsilon e^{\lambda_\textsc{q} t}  ,
	\end{equation}
	where the exponential behavior holds only in systems with scrambling over the appropriate period $t_s \ll t \ll t_\ts{e}$. 	{\aurelia  The SOV-OTOC connection is complementary to the optimal-path approach to study chaos in continuously monitored systems \cite{Lewalle2018}.}
 Note that the SOV $\Delta \ha A_t^2$ 
	is an \pma{observable} constructed from the knowledge of the evolution under different noise realizations. 
	The classical limit of the above equation is similar, the only difference being that $\Delta A_t^2$ becomes a function of time and the trace an average over a region of phase space \cite{wang_quantum_2021}. 

	The short-time decay  of the OTOC, $C_t \sim C_0 e^{- t/\tau_D}$, is characterized by the dissipation time $\tau_D = \big(2 \gamma \Tr([\ha L, [\ha L, \ha A]]^2)/(C_0 N)\big)^{-1}$ and the initial value $C_0=\Tr([\ha L, \ha A]^2)/N$. Interestingly, this dissipation time is related to the Hilbert-Schmidt norm of the dissipator acting on the initial operator (see App. \ref{sec2}). {\pablo It is analogous to the decoherence rate found in \cite{xu_extreme_2019} but now obtained for operators in the Heisenberg picture.
		When $[\ha H_0, \ha L]=0$, the operators share a common eigenbasis, and using $\ha L \ket{n} = l_n\ket{n}$, we can write the dissipative OTOC \eqref{eq:otoc} as 
		\begin{equation} \label{dissOTOC}
			C_t = \sum_{m,n} (l_m - l_n)^2 e^{-2 \gamma (l_m - l_n)^2 t}|A_{nm}|^2,
		\end{equation}
		where $A_{nm} = \braket{n|\ha A|m}$. From this expression,  the two exponentially decaying regimes reported in \cite{syzranov_out--time-order_2018} appear from the largest and smallest eigenvalue differences governing the short- and long-time dynamics, respectively.}
	
	{\pablo Note that, for simplicity, we have focused on the infinite temperature OTOC. However, the connection to OTOC \eqref{eq:otoc} can be generalized to 
		unregularized thermal  OTOC \cite{tsuji_out--time-order_2018}  by tracing \eqref{eqVar} over a thermal state $\rho_\beta = e^{-\beta \ha H}/\Tr(e^{-\beta \ha H})$. In this case, the SOV-OTOC relation also involves an additional expectation value of the Lindbladian $\Tr(\mc L[\Delta \ha A_t^2]\rho_\beta)$, which is non-zero in general. Fidelity OTOCs \cite{lewis-swan_unifying_2019} can be obtained by taking the jump operator to be a projector over a pure state. 		Regularized OTOCs \cite{maldacena_bound_2016} can also be obtained by modifying the jump operator as $\ha L \rightarrow \rho_\beta^{1/4} \ha L \rho_\beta^{1/4}$ and tracing over the completely mixed state.}

	We next illustrate our findings in the Lipkin-Meshkov-Glick (LMG) model subject to energy dephasing and characterize its Lyapunov exponent using the SOV-OTOC \pma{connection}.

	\begin{figure}
		\includegraphics{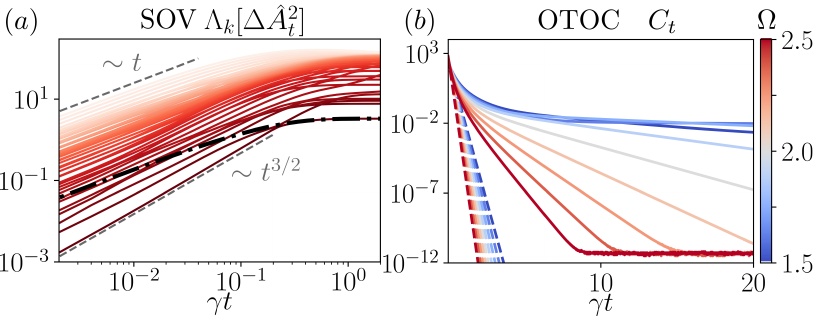}
		\caption{\textbf{Evolution of  (a) \pma{the SOV eigenvalues}, and (b) the OTOC $C_t$ for the quantum sLMG model}. The  operator $\ha A=(\ha S_x + \ha S_y + \ha S_z)/\sqrt{3}$ evolves under the stochastic Hamiltonian \eqref{eq:HsLMG} with $\gamma = 2$, $\Omega = 1$, and $S = 20$. (a) \pma{Eigenvalues of the SOV as a function of time (solid red) and expectation value of the SOV for the state which minimized the deviation at long time, $\braket{\Psi|\Delta\ha A^2_t|\Psi}$ (black dash-dotted).} (b) Dissipative OTOC obtained from the SOV-OTOC relation  \eqref{eq:otoc} (solid line) and short-time expansion (dashed line) \pma{for different values of $\Omega$ (colorbar)} across the phase transition---at $\Omega_c=2$. 
		}
		\label{fig:qsLMG_dissOTOC}
	\end{figure}

	\emph{Stochastic Lipkin-Meshkov-Glick (sLMG) model}.---   The LMG model describes the collective motion of $N$ identical two-level systems  fully connected to each other with the same coupling strength \cite{lipkin_validity_1965}. Its quantum Hamiltonian reads 
	\beq \label{eq:HLMG}
	\ha H_\textsc{lmg} = \Omega \ha S_z - \frac{2}{N} \ha S_x^2, 
	\eeq
	where $\Omega$ is the frequency in units of the coupling strength, and $\ha S_j$ are the general spin operators of dimension $2S+1$. We stay in the sector $S = N/2$. Since the total spin $\ha{\be S}^2 = \ha S_x^2 + \ha S_y^2 + \ha S_z^2$ commutes with the spin operators, $[\ha S_j, \ha{\be S}^2]=0$,  the total angular momentum is conserved. 
	Due to time-translational symmetry, energy is conserved, and since there is only one degree of freedom, this model is integrable. If  this continuous symmetry is broken by  periodic kicks in $\ha S_x^2$, the model turns into the known kicked top \cite{haake_quantum_2010}. 
	
	Here, we break time-translational symmetry by adding  noise in the energy scale and consider   
	\beq \label{eq:HsLMG}
	\ha H_t = \ha H_\ts{lmg} (1 + \sqrt{2 \gamma} \xi_t). 
	\eeq
	This leads to dephasing in the energy eigenbasis at the ensemble level. \pma{The evolution of the SOV  instantaneous (ordered) eigenvalues satisfying $\Delta \ha A_t^2 \ket{v_k(t)} = \Lambda_k(t)\ket{v_k(t)}$, for this model is shown in Fig. \ref{fig:qsLMG_dissOTOC}(a).
		In analogy with the theory of quantum transport \cite{jayannavar_nondiffusive_1982, hislop_transport_2019}, we find \textit{diffusive modes} in which $\Lambda_k(t) \sim t$ and \textit{superdiffusive modes} in which $\Lambda_k(t) \sim t^{3/2}$. In the more general case $[\ha H_0, \ha L]\neq 0$, one also finds ballistic modes $\Lambda_k(t) \sim t^2$, as we detail in  App. \ref{sec4}. Based on the eigenvalues, one can find a state which minimizes the SOV at long times
		\begin{equation} \label{minSOV}
			\ket{\Psi} = \lim_{t \rightarrow \infty} \ket{v_0(t)},
		\end{equation}
		which corresponds to the steady state of the eigenvector with the smallest deviation, and is thus the state minimally affected by the noise. The expectation value over this state is shown in Fig. \ref{fig:qsLMG_dissOTOC}(a) (black dash-dotted) for the sLMG model, and we verify that it minimizes the spread at long times.}
	Fig. \ref{fig:qsLMG_dissOTOC}(b) shows the evolution of the dissipative OTOC, computed from the SOV using Eq.~\eqref{eqExplambdat}. 
		\pma{We observe several exponential decays, as apparent from \eqref{dissOTOC} and in agreement with \cite{syzranov_out--time-order_2018}.}
	This illustrates how the SOV can be used to obtain the dissipative OTOC.

	The classical limit of the Hamiltonian \eqref{eq:HLMG} is obtained by taking its expectation value over \pma{SU(2) coherent states $\ket{\zeta}$ \cite{perelomov_coherent_1986} (see App. \ref{sec5})} 
	in the thermodynamic limit, $N \rightarrow \infty$.  We introduce the canonical variables $Q$ and $P$ as 
	$\zeta = \frac{Q - i P}{\sqrt{4 - (Q^2 + P^2)}}$ \cite{pilatowsky-cameo_positive_2020, pappalardi_scrambling_2018}, which yields $H_\ts{lmg} = \lim_{S \rightarrow \infty} \braket{\zeta|\ha H_\ts{lmg}|\zeta} /S = \tfrac{\Omega}{2}P^2 + \left(\tfrac{\Omega}{2}-1\right) Q^2 + \tfrac{1}{4}(Q^2 P^2 + Q^4),$ 
	where the terms of $\mc O(1/N)$ are neglected (see App. \ref{sec5}).
	This model is integrable and exhibits an unstable fixed point at the origin, $Q^* = P^* = 0$, for $0 < \Omega < 2$. Since scrambling originates from an unstable point, it is already present in the semiclassical limit \cite{xu_does_2020, pilatowsky-cameo_positive_2020, pappalardi_scrambling_2018}.

	Here, we consider the classical equivalent of \eqref{eq:HsLMG}, namely, $H_t = H_\ts{lmg}(1 + \sqrt{2 \gamma} \xi_t)$.  The evolution of the noise-averaged observable displays the classical analog of energy dephasing, namely $\partial_t\av{A_t} = - \pb{H_\ts{lmg}, \av{A_t}} + 2 \gamma \pb{H_\ts{lmg}, \pb{H_\ts{lmg}, \av{A_t}}}$, where $\pb{f, g}$ denotes the Poisson bracket of $f$ and $g$. 
	We characterize the Lyapunov exponent using three complementary methods: 
	Analytically, the approach proposed by van Kampen \cite{kampen_stochastic_1992, van_kampen_stochastic_1976} yields the Lyapunov for the sLMG as (see App. \ref{sec5})
	$\lambda^{(1)} = \sqrt{2 \Omega - \Omega^2} - \gamma (2 \Omega -  \Omega^2).$
	(ii) Second, numerically. The standard, classical definition of the average Lyapunov exponent  gives 
	{\pablo 
		$\lambda^{(2)} = \av{\lim_{t \rightarrow \infty} \ln (\ell_t/\ell_0)/t},$
		where $\ell_t^2 = (Q_t {-} Q_t')^2 + (P_t{-} P_t')^2$ is the distance between two initially close trajectories evolving with the same realization of the noise, which is found solving the stochastic Hamilton's equations with a stochastic Runge-Kutta method \cite{kloeden_numerical_1992} (see App. \ref{sec6}). 
		(iii) Finally, our formalism gives the Lyapunov from Eq. \eqref{eq:otoc}, by taking as observable the position $A_t=Q_t$,  the jump operator being the Hamiltonian itself,  $L = H_0$. 
		Namely, 
		$\lambda^{(3)} = \lim_{t \rightarrow \infty} \frac{1}{2t}\ln \left( \mr d_t \Delta Q_t^2/\epsilon \right)$.}

	\begin{figure}
		\includegraphics{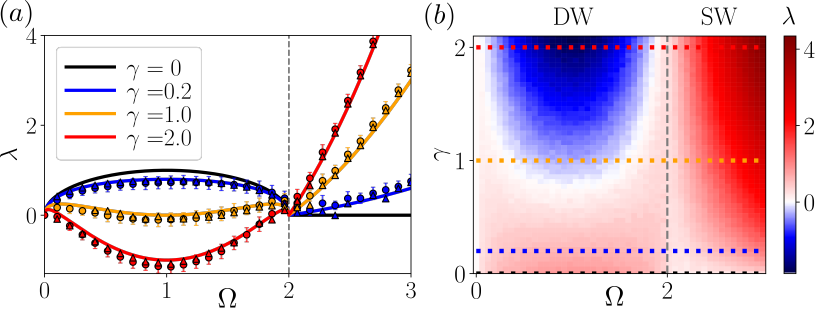}
		\caption{ \textbf{Lyapunov exponent of the classical sLMG model at the saddle point} $Q^* = P^* = 0$ as function of $\Omega$ (a) for different values of the noise strength $\gamma$ and (b) over the  phase diagram. (a) $\lambda$ as computed (i) analytically using van Kampen's method $\lambda^{(1)}$ (solid lines), (ii) from the standard definition  $\lambda^{(2)}$(circles with errorbar), and (iii) from the SOV-OTOC connection $\lambda^{(3)}$ (triangles). The known results for the  LMG correspond to $\gamma=0$ (black). (b) \textbf{Phase diagram}. The color scale represents the Lyapunov exponent $\lambda^{(2)}$ as a function of the model parameter $\Omega$ and the noise strength $\gamma$. A positive value of $\lambda$ (red) implies exponential divergence of close initial conditions, while a negative value (blue) indicates exponential convergence. The dotted horizontal lines represent the values of $\gamma$ sampled in (a). The vertical dashed gray line represents the transition between the double well ($\Omega < 2$) and single well  ($\Omega \geq 2$) phase.}
		\label{fig:Lyap_sLMG}
	\end{figure}
	
	Figure \ref{fig:Lyap_sLMG}(a) shows the Lyapunov exponent $\lambda$ obtained from the above methods as a function of $\Omega$ for  different noise strengths. We verify that the three methods are in good agreement up to numerical errors. \pma{For $\gamma=0$  (black line), we recover the behavior of LMG with $\lambda >0$ in the double well (DW) phase and $\lambda=0 $ in the single well (SW) phase.} Introducing a weak stochastic perturbation with $\gamma$, the Lyapunov exponent becomes smaller in the DW phase---trajectories diverge more slowly---while the SW phase acquires a positive $\lambda$. Increasing the level of noise causes the Lyapunov exponent in the DW phase to decrease and even reach  negative values,
	where trajectories converge exponentially, while the value of $\lambda$ in the SW phase increases. 
	This rich behavior is summarized in the phase diagram presented in Fig. \ref{fig:Lyap_sLMG}(b). The latter further shows that, under the application of strong noise, the origin $(Q^*, P^*)=(0,0)$ is stable in the DW phase---trajectories converge exponentially to it---while it is unstable in the SW phase---trajectories diverge exponentially from it. Therefore the sLMG in the DW phase shows a noise-induced transition to stability \cite{horsthemke_noise-induced_2006}, analogous to the stabilization seen for periodic driving \cite{defenu2023outofequilibrium}.

	In summary, we have introduced the stochastic operator variance \pma{and shown it is a valuable tool to study quantum systems driven by noise. We have shown that this observable obeys an uncertainty relation \eqref{SOV_unc_rel}, and that it can be used to identify the state \eqref{minSOV} which minimizes deviation from the Lindblad dissipative dynamics. We have provided a bipartite interpretation of the SOV. In addition, our results} unveil a SOV-OTOC \pma{connection \eqref{eqExplambdat}}, which provides an operational protocol harnessing noise as a resource to probe OTOC and extract the Lyapunov exponent in noisy quantum chaotic systems. To illustrate our results, we introduced a stochastic generalization of  LMG model and characterized its behavior in the quantum and classical realms. Our results provide the means to elucidate the fate of quantum chaos in noisy systems and benchmark NISQ devices.

	{\it Acknowledgements.---} We thank Niklas H\"ornedal, \pma{Howard Wiseman, Andrew Jordan}, Federico Roccati,  Federico Balducci, and Ruth Shir for insightful discussions and comments on the manuscript. This research was partly funded by the Luxembourg National Research Fund (FNR), Attract grant 15382998, and the John Templeton Foundation (Grant 62171).  The opinions expressed in this publication are those of the authors and do not necessarily reflect the views of the John Templeton Foundation. For the purpose of open access, the authors have applied a Creative Commons Attribution 4.0 International (CC BY 4.0) license to any Author Accepted Manuscript version arising from this submission.


\let\oldaddcontentsline\addcontentsline
\renewcommand{\addcontentsline}[3]{}
\let\addcontentsline\oldaddcontentsline

\newpage
\appendix
\onecolumngrid
\newpage

\section{Stochastic evolution: details on the main derivation} \label{sec1}
\subsection{On the Heisenberg picture of an explicitly time-dependent Hamiltonian}

We consider an explicitly time-dependent Hamiltonian $\ha H_t$, which in general does not commute with itself at different times, $[\ha H_t, \ha H_{t'}]\neq 0$. It governs the dynamics of a state through  the Schr\"odinger equation 
\begin{equation}
	\frac{\mr d}{\mr dt}\ket{\psi_t} = - i \ha H_t \ket{\psi_t},
\end{equation}
whose solution $\ket{\psi_t}= \ha U_t \ket{\psi_0} = \mc T_{\leftarrow} e^{- i \int_0^t \ha H_\tau \mr d \tau}\ket{\psi_0},$ involves the propagator from time $0$ to $t$, $\ha U_t$, where $\mc T_{\leftarrow}$ is the chronological time-ordering operator \cite{breuer_theory_2002}. An (explicitly time-independent) operator $\ha A$ in the Heisenberg picture then evolves as
\begin{equation}\label{evolAt_heis}
	\ha A^\textsc{(h)}_t = \ha U_t\dg \ha A \ha U_t.
\end{equation}
Differentiating this evolution gives Heisenberg's equation for the evolution of $\ha A_t$ as
\begin{align} \label{Heis_eq}
	\frac{\mr d}{\mr dt}\ha A^\textsc{(h)}_t &= \dot{\ha U}_t\dg \ha A \ha U_t + \ha U_t\dg \ha A \dot{\ha U}_t 
	= i \ha U_t\dg \ha H_t \ha A \ha U_t - i \ha U_t\dg \ha A \ha H_t \ha U_t
	= i [\ha H^\textsc{(h)}_t, \ha A^\textsc{(h)}_t ],
\end{align}
where $\ha H^\textsc{(h)}_t = \ha U_t\dg \ha H_t \ha U_t \neq \ha H_t$ is the Hamiltonian in the Heisenberg picture with respect to itself, which is not equal to $\ha H_t$ since it does not commute with itself at the different times $t' < t$ included in $\ha U_t$. Note that the Heisenberg equation \eqref{Heis_eq} is recovered when we introduce the Hamiltonian in the Heisenberg picture with respect to itself $\ha H^\textsc{(h)}_t$, but that its solution \eqref{evolAt_heis} only requires the bare Hamiltonian $\ha H_t$ in the propagator $\ha U_t$.

We bring attention to the fact that Heisenberg operators evolve \textit{``backwards in time"}. This can be illustrated by splitting the evolution over two different times $0\leq t_1 \leq t_2 $. The expectation value of any general operator $\ha A$ then reads
\begin{equation}
	\braket{\psi| \mc T_{\rightarrow} e^{+i \int_{0}^{t_1} \ha H_\tau \mr d \tau } \mc T_{\rightarrow} e^{+i \int_{t_1}^{t_2} \ha H_\tau \mr d \tau } \ha A \mc T_{\leftarrow} e^{-i \int_{t_1}^{t_2} \ha H_\tau \mr d \tau }\mc T_{\leftarrow} e^{-i \int_{0}^{t_1} \ha H_\tau \mr d \tau }|\psi},
\end{equation}
where $\mc T_{\rightarrow}$ denotes anti-chronological time-ordering.
While states evolve forwards in time, $0 \rightarrow t_1 \rightarrow t_2$, the equivalent evolution at the level of operators is for them to evolve backwards in time, $t_2 \rightarrow t_1 \rightarrow 0$. Of course this is just a result of the Heisenberg representation and does not correspond to a backwards time evolution.

\subsection{Heisenberg evolution with a stochastic Hamiltonian}

Adjoint stochastic master equations (SMEs) have been studied in \cite{gammelmark_past_2013, warszawski_solving_2020} for continuous homodyne detection and read
\begin{equation}\label{SME_homodyne}
	\mr d \ha A_t = i [\ha H, \ha A] \mr d t + \sqrt{\eta} (\ha c\dg \ha A_t + \ha A_t \ha c) \mr d Y_{t-\mr dt} + \sum_m \left(\ha L_m \ha A_t \ha L_m - \frac{1}{2}\{\ha L_m\dg \ha L_m, \ha A_t\}\right) \mr dt,
\end{equation}
where we use the backwards differential $\mr d \ha A_t \equiv \ha A_{t-\mr dt}- \ha A_t$. $\eta$ is the efficiency of the detector, $\ha c$ is the measurement observable, $\ha L_m$ are the Lindblad operators describing coupling to a general bath, and $\mr d Y_t$ is the measurement record. $\mr d Y_t$ is a stochastic process.  
Note that this equation does not preserve trace. When $\mr d Y_t$ represents white noise without drift, i.e. $\av{\mr d Y_t}=0$, the average over the noise recovers the standard adjoint master equation for the average operator $\av{\ha A_t}$. 

We draw attention to the fact that the measurement record appears as $\mr d Y_{t-\mr dt}$. To apply It\=o's rules, we need to evaluate the noise at the beginning of the time interval. In the Schr\"odinger picture, this corresponds to evaluating the noise at time $t$ in the interval $[t, t+\mr dt]$. But since we work in the Heisenberg picture we propagate the operator in the interval $[t-\mr dt, t]$ and we have to evaluate the noise at $t-\mr dt$.
The propagator over $\mr dt$ thus reads
\begin{equation}
	\ha U_{dt} = e^{-i \ha H \mr d t - i \sqrt{2 \gamma} \ha L \mr d W_{t - \mr dt}},
\end{equation}
which gives the SME \eqref{eomAt}. Note that this equation is trace preserving. So although in the case of driftless white noise $\av{\mr d Y_t}=0$, the SME's \eqref{eomAt} and \eqref{SME_homodyne} give the same results at the average level, they describe different dynamics at the level of single trajectories.

\subsection{Finding the equation of motion for the SOV}

The equation of motion for the second stochastic moment simply reads  $\frac{d}{dt}\av{\ha A^2_t}= \mc L\dg[\av{\ha A^2_t}]$, and that for the first moment squared reads
\begin{equation}
	\frac{d}{dt}\av{\ha A_t}^2= \mc L\dg[\av{\ha A_t}]\av{\ha A_t}+ \av{\ha A_t}\mc L\dg[\av{\ha A_t}].
\end{equation}
In the latter, we want to introduce the term $\mc L\dg[\av{\ha A_t}^2]$ in order to recover $\mc L\dg[\Delta \ha A^2_t]$ through linearity of the Lindbladian superoperator. The difference reads
\begin{align} \notag
	\frac{\mr d}{\mr dt}\av{\ha A_t}^2 - \mc L\dg[\av{\ha A_t}^2] &= -\gamma \left( \left\{[\ha L, [\ha L, \av{\ha A_t}]] ,\av{\ha A_t}\right\}  - [\ha L, [\ha L, \av{\ha A_t}^2]]\right) \\ \notag
	&= -2 \gamma \bigg(\ha L \av{\ha A_t}^2 \ha L + \av{\ha A_t} \ha L^2 \av{\ha A_t} 
	- \ha L \av{\ha A_t} \ha L\av{\ha A_t} - \av{\ha A_t} \ha L\av{\ha A_t} \ha L \bigg) \\
	&= - 2 \gamma [\ha L, \av{\ha A_t}]^2, 
\end{align}
which yields the equation of motion \eqref{eqVar} for the SOV given in the main text
\begin{equation}
	\frac{\mr d}{\mr dt} \Delta \ha A_t^2 = \mc L\dg[\Delta \ha A_t^2] - 2 \gamma [\ha L, \av{\ha A_t}]^2.
\end{equation}

\subsection{The SOV uncertainty principle}

The theory of positive matrices (see pg. 75 \cite{bhatia_positive_2009}) tells us that given a completely positive unital map, like $e^{\mc L\dg t}[\bullet]$, for all operators $\ha A, \; \ha B$, the following matrix is positive semidefinite
\begin{equation}
	\left(\begin{array}{cc}
		\Delta \ha A_t^2 & \Delta \widehat{A B}_t \\
		\Delta \widehat{A B}_t\dg & \Delta \ha B^2_t
	\end{array}\right) \geq 0,
\end{equation}
i.e. the $2 N \times 2 N$ matrix has to be positive semidefinite. While this statement is powerful, it does not give any straightforward inequality. For this reason, we take the expectation value of each of the blocks over a certain state $\rho$, denoted $\avr{\bullet}=\Tr(\bullet \rho)$. This allows recasting the inequality for a 2x2 matrix, 
\begin{equation}
	\left(\begin{array}{cc}
		\avr{\Delta \ha A_t^2} & \avr{\Delta \widehat{A B}_t} \\
		\avr{\Delta \widehat{A B}_t\dg} & \avr{\Delta \ha B^2_t}
	\end{array}\right) \geq 0.
\end{equation}
Let us now apply Sylvester's criterion: a matrix is positive semi-definite if and only if all its principal minors, i.e. the determinants of the matrix in which we delete the columns and rows with the same index, are non-negative.  Doing so, we can recast the complete positivity condition into
\begin{equation}
	\avr{\Delta \ha A_t^2}\geq 0, \quad \avr{\Delta \ha B^2_t}\geq 0, \quad \avr{\Delta \ha A_t^2}\avr{\Delta \ha B_t^2} - |\avr{\Delta \widehat{A B}_t}|^2 \geq 0.
\end{equation}
The two first inequalities above are fulfilled because the SOV is positive semidefinite, the last condition yields an analog of the Schwarz inequality 
\begin{equation}
	\avr{\Delta \ha A_t^2}\avr{\Delta \ha B_t^2} \geq |\avr{\Delta \widehat{A B}_t}|^2.
\end{equation}
Let us now expand the r.h.s. of the inequality. Writing $\ha A \ha B = \frac{1}{2}(\{\ha A, \ha B\} + [\ha A, \ha B])$ and $ \ha B \ha A = \frac{1}{2}(\{\ha A, \ha B\} - [\ha A, \ha B])$, we obtain
\begin{align*}
	|\avr{\Delta \widehat{A B}_t}|^2 &= (\avr{e^{\mc L\dg t}[\ha A \ha B]}-\avr{\ha A_t \ha B_t})(\avr{e^{\mc L\dg t}[\ha B \ha A]}-\avr{\ha B_t \ha A_t}) \\ &= \frac{1}{4}\bigg( \avr{e^{\mc L\dg t}[\{\ha A, \ha B\}]}^2 - \avr{e^{\mc L\dg t}[[\ha A, \ha B]]}^2 - 2 \avr{e^{\mc L\dg t}[\{\ha A, \ha B\}]} \avr{\{\ha A_t, \ha B_t\}} + 2 \avr{e^{\mc L\dg t}[[\ha A, \ha B]]}\avr{[\ha A_t, \ha B_t]} \\ & \hspace{4mm} + \avr{\{\ha A_t, \ha B_t\}} - \avr{[\ha A_t, \ha B_t]}^2\bigg) \\
	&= \frac{1}{4}\left( \avr{e^{\mc L\dg t}[\{\ha A, \ha B\}]} - \avr{\{\ha A_t, \ha B_t\}}\right)^2 - \frac{1}{4}\left( \avr{e^{\mc L\dg t}[[\ha A, \ha B]]} - \avr{[\ha A_t, \ha B_t]}\right)^2\\
	&=\frac{1}{4}\left(D_+^2(\ha A, \ha B) - D_-^2(\ha A, \ha B)\right),
\end{align*}
In the last line, we introduced $D_\eta(\ha A, \ha B)=\avr{e^{\mc L\dg t}([\ha A, \ha B]_\eta)} - \avr{[\ha A_t, \ha B_t]_\eta}$, where  $ [\ha X, \ha Y]_\eta= \ha X \ha Y + \eta \ha Y \ha X $ with $\eta = \pm 1$ is the generalized commutator. Therefore the SOV uncertainty principle reads
\begin{equation}
	\Tr({\Delta \ha A_t^2}\rho)\Tr({\Delta \ha B_t^2}\rho) \geq \frac{1}{4}\left(D_+^2(\ha A, \ha B) - D_-^2(\ha A, \ha B)\right).
\end{equation}

\section{Dissipative out-of-time-order correlators} \label{sec2}
\subsection{Short-time expansion of OTOC}
\begin{figure}[ht]
	\centering
	\includegraphics[width = .5 \linewidth]{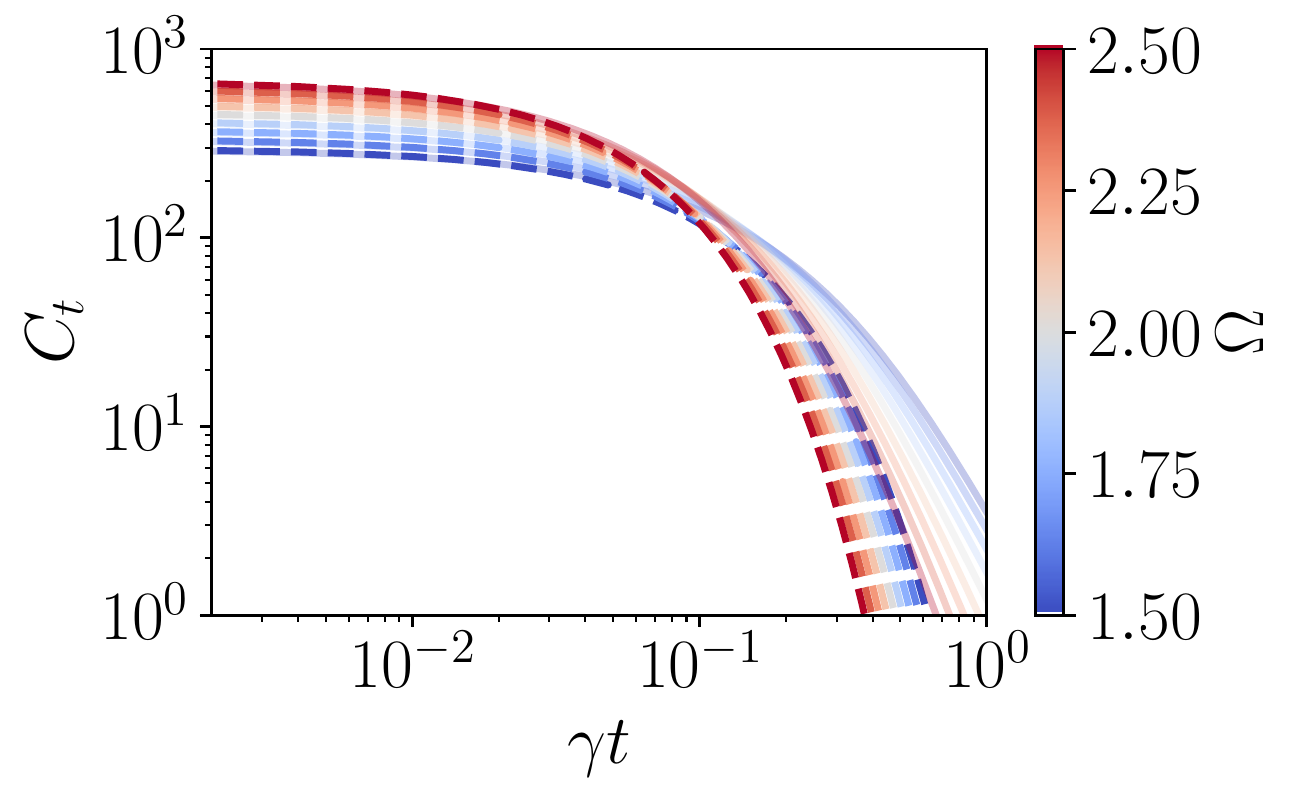}
	\caption{\textbf{Short-time behavior of the OTOC as computed from the SOV-OTOC relation}. Full OTOC (solid line) and short-time expansion \eqref{eq:shortTimeOTOC} (dashed line). The crossover between the two exponentially decaying regimes is seen around $\gamma t \approx 0.1$. }
	\label{fig:shortTime_OTOC}
\end{figure}

It is insightful to consider the short-time expansion of the OTOC 
\begin{align}
	\mr d_t &\Tr(\Delta \ha A_t^2) = - 2 \gamma \Tr\left([\ha L, \ha A + t \mc L\dg[\ha A]+\mc O(t^2)]^2\right) 
	= - 2 \gamma \Tr\left([\ha L, \ha A]^2 + 2 t [\ha L, \ha A][\ha L, \mc L\dg[\ha A]]\right) + \mc O(t^2). 
\end{align}
The linear term in $t$ has an oscillatory contribution $-4i \gamma t \Tr([\ha L, \ha A][\ha L, [\ha H_0, \ha A]])$ and a dissipative contribution
\begin{align}
	-C_0 \frac{t}{\tau_D} &= \frac{2 \gamma}{N} t \Tr([\ha L, \ha A][\ha L, [\ha L, [\ha L, \ha A]]]) 
	= - \frac{2 \gamma}{N} t \Tr([\ha L, [\ha L, \ha A]]^2),
\end{align}
where 
we used the cyclic property of the trace to write $\Tr(\ha X [\ha Y, \ha Z])= \Tr([\ha X, \ha Y] \ha Z)$ with $\ha X = [\ha L, \ha A], \; \ha Y = \ha L, $ and $ \ha Z = [\ha L, [\ha L, \ha A]]$. Then the inverse dissipation time, $\tau_D^{-1} \propto  2 \gamma \Tr(\mc D[\ha A]^2) = 4 (\mc D[\ha A], \mc D[\ha A])\geq 0$ is related to the Hilbert-Schmidt norm of the dissipator $\mc D[\bullet]=- [\ha L, [\ha L, \bullet]]$ acting on the initial operator $\ha A$. The short-time behavior is then determined by the exponential
\begin{equation}\label{eq:shortTimeOTOC}
	C_t \approx C_0 e^{-t/\tau_D}.
\end{equation}
This approximation is compared to the full OTOC in Fig. \ref{fig:shortTime_OTOC}, where the crossover between the different exponential regimes of the dissipative OTOC is apparent.

\subsection{The dissipative OTOC for $[\ha H_0, \ha L]=0$}

In the simple case that $[\ha H_0, \ha L]=0$, the operators $\ha H_0$ and $\ha L$ share a common eigenbasis, such that $\ha H_0 = \sum_n E_n \ket{n}\bra{n} $ and $ \ha L = \sum_n l_n \ket{n}\bra{n}$. The solution of the adjoint master equation then reads
\begin{equation}
	\ha A_t= \sum_{m,n} A_{mn} e^{i (E_m - E_n)t - \gamma (l_m - l_n)^2 t} \ket{m}\bra{n}.
\end{equation}
The behavior of the dissipative OTOC then is simply
\begin{align}
	\Tr([\ha L, \ha A_t]^2) &= \sum_{m,k} (l_m - l_k)^2 |A_{km}(t)|^2 
	= \sum_{m,k} (l_m - l_k)^2 e^{-2 \gamma (l_m - l_k)^2 t}|A_{km}|^2. 
\end{align}
This shows that the short-time dynamics is governed by the two eigenvalues with largest difference $\max_{m,k}(l_m-l_k)^2$ over which the operator has support, i.e. $A_{mk}\neq 0$, and the long-time dynamics is governed by the smallest possible non-zero difference $\min_{m,k}(l_m - l_k)\neq 0$ over which the initial operator has support.  

\section{Product of operators as a doubled Hilbert space operation} \label{sec3}

The standard product between operators in a Hilbert space $\mathscr H$ with closure relation $\ha{\mbb 1}_{\msc H} = \sum_n \ket{n}\bra{n} $ reads 
\begin{equation}
	\ha X  \ha Y = \sum_{n,m,k} \braket{n|\ha X|k}\braket{k|\ha Y|m}\ket{n}\bra{m}.
\end{equation}
In turn, the tensor product of the operators over a doubled Hilbert space $\msc H = \msc H_1 \otimes \msc H_2$ with closure relation $\mbb 1 = \mbb 1_{\msc H_1}\otimes \mbb 1_{\msc H_2}= \sum_{n_1, n_2}\ket{n_1, n_2}\bra{n_1, n_2}$ where $\ket{n_1, n_2}= \ket{n_1}\otimes \ket{n_2}$ reads
\begin{equation}
	\ha X \otimes \ha Y = \sum_{n_1, m_1, n_2, m_2} \braket{n_1|\ha X|m_1}\braket{n_2|\ha Y|m_2} \ket{n_1, n_2}\bra{m_1, m_2}.
\end{equation}
Thus, applying the swap operator, defined from  $\mbb S \ket{n_1}\otimes\ket{n_2} = \ket{n_2}\otimes\ket{n_1}$, yields
\begin{equation}
	(\ha X \otimes \ha Y)\mbb S = \sum_{n_1, m_1, n_2, m_2} \braket{n_1|\ha X|m_1}\braket{n_2|\ha Y|m_2} \ket{n_1, n_2}\bra{m_2, m_1}.
\end{equation}
The partial trace over the second Hilbert space $\Tr_{\msc H_2}(\bullet) = \sum_{n_2}\braket{n_2|\bullet|n_2}$ gives
\begin{align} \notag
	\Tr_{\msc H_2}\left((\ha X \otimes \ha Y)\mbb S\right) &= \sum_{n_1, m_1, m_2} \braket{n_1|\ha X|m_1}\braket{m_1|\ha Y|m_2} \ket{n_1}\bra{m_2}
	= \ha X  \ha Y.
\end{align}

\section{Quantum stochastic Lipkin Meshkov Glick model} \label{sec4}
\subsection{Transport in quantum sLMG}
\begin{figure}[ht]
	\centering
	\includegraphics[width = .5\linewidth]{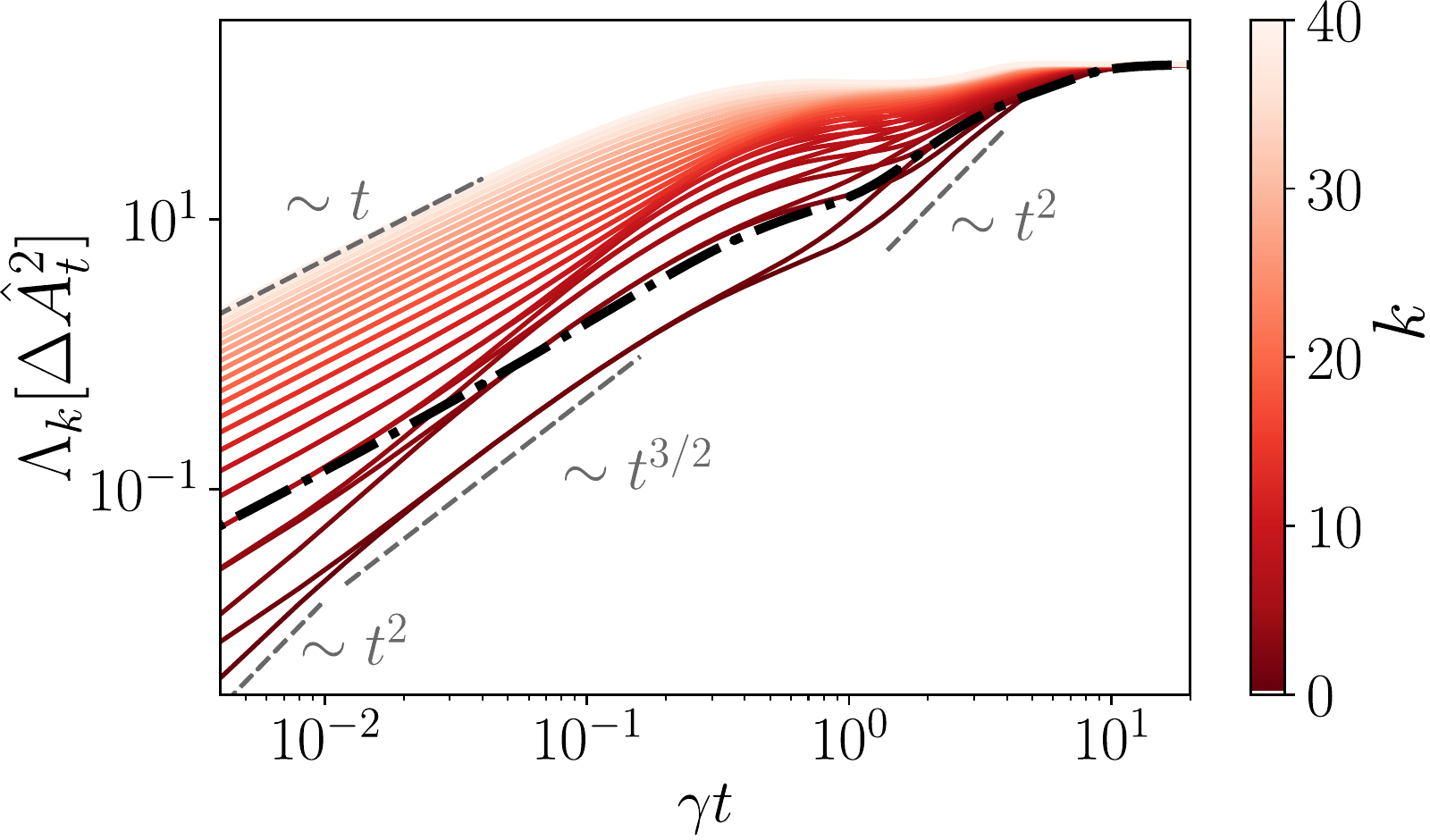}
	\caption{Eigenvalues of the SOV as a function of time for  noncommuting Hamiltonian and Jump operator $[\ha H_0, \ha L]\neq 0$ (red, colorscale shows order of the eigenvalues $k$) and the expectation value over the minimum SOV state $\braket{\Psi|\Delta \ha A_t^2|\Psi}$ (black dash-dotted). The parameters are $\gamma =2, \Omega = 1, S = 20, \ha A = (\ha S_x + \ha S_y + \ha S_z)/\sqrt{3}$  and $\ha L = \ha S_x$. }
	\label{fig:evalsSOV}
\end{figure}
In the main text, we presented the eigenvalues of the SOV and their transport analog in the sLMG for $[\ha H_0, \ha L]=0$, see Fig. \ref{fig:qsLMG_dissOTOC} (a). The case in which $[\ha H_0, \ha L] \neq 0$ shows a richer phenomenology, even for the simple sLMG model. The eigenvalues of the SOV are shown in Fig. \ref{fig:evalsSOV}. The smallest eigenvalue first evolves ballistically,  $\Lambda_0(t) \sim t^{2}$, then turns into superdiffusive $\Lambda_0(t) \sim t^{3/2}$ before changing further to a slower growth $t^{\alpha}$ with $\alpha<1$, to ballistic again $\Lambda_0(t) \sim t^{2}$ and then saturates. Another interesting feature is that at long times all eigenvalues saturate to the same value, a feature not present in the case $[\ha H_0, \ha L] =0$.

\subsection{Visualizing the quantum stochastic operator variance}

The SOV  $\Delta \ha A_t^2$ that we introduced is an operator over the Hilbert space. We illustrate this quantity for the case of the quantum sLMG model. In general, this operator is $d$ dimensional, and it has non-zero components over all the elements of the considered operator basis. However, since the sLMG is a mean-field description, we are mainly interested in its projection over the subspace spanned by $\{\mbb 1, \ha S_x, \ha S_y, \ha S_z\}$. We would like to have these elements be orthonormal with each other. For this reason, we choose the Hilbert-Schmidt inner product
\beq
( \ha A, \ha B ) =\frac{1}{S(S+1)(2S+1)/3} \Tr(\ha A\dg \ha B), 
\eeq
where the normalization has been chosen such that $(\ha S_i, \ha S_j)=\delta_{ij}$ and  comes from $\Tr(\ha S_z^2) = \sum_{j=0}^{2 S} (S-j)^2= S(S+1)(2S+1)/3.$ Note that this normalization leads to $(\mbb 1, \mbb 1)=\frac{3}{S(S+1)}\neq 1$.

\begin{figure*}[ht]
	\includegraphics[width = .25 \linewidth]{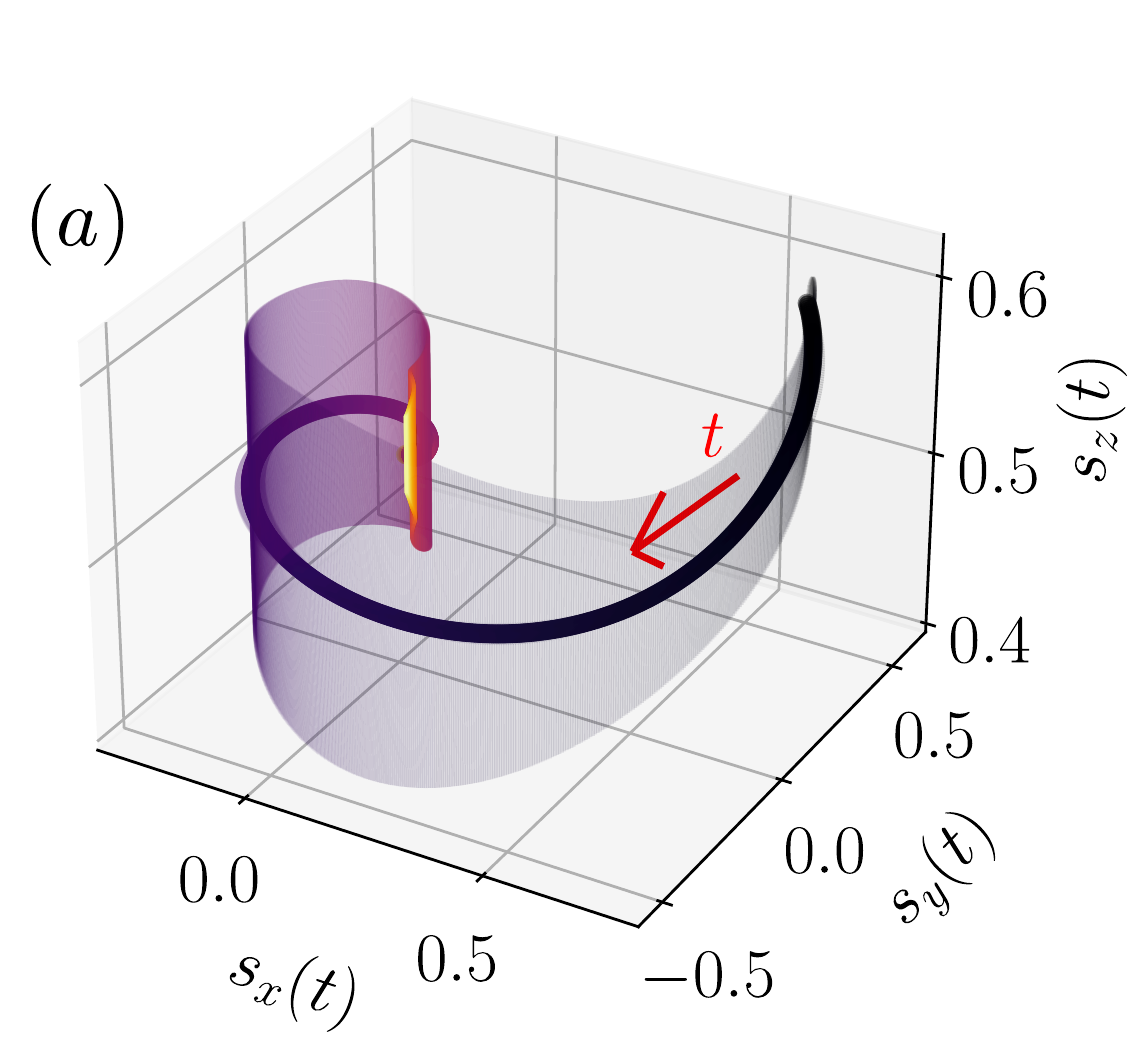}
	\includegraphics[width = .74 \linewidth]{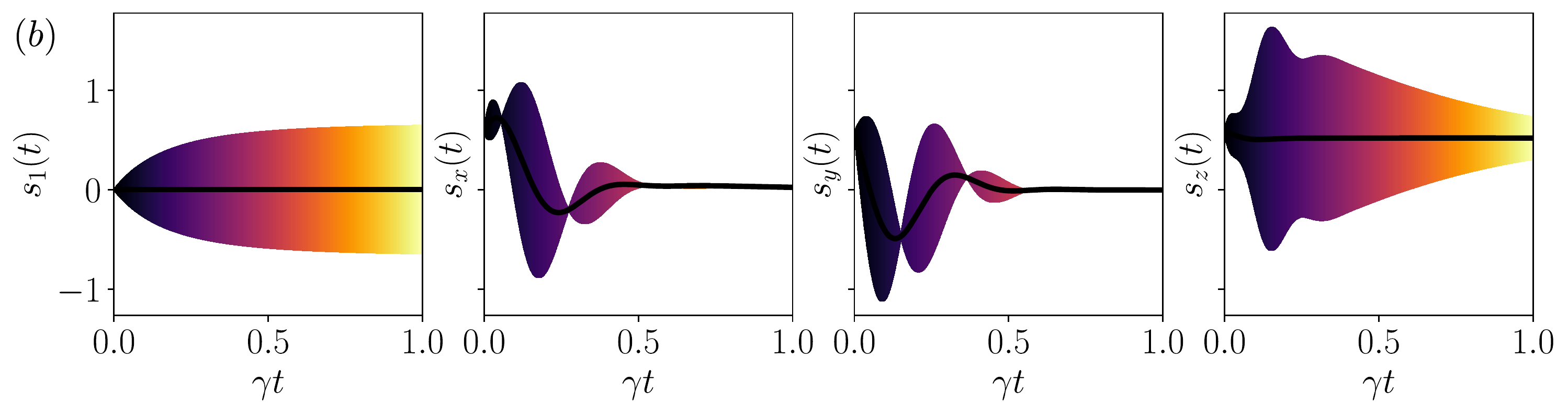}
	\includegraphics[width = .25 \linewidth]{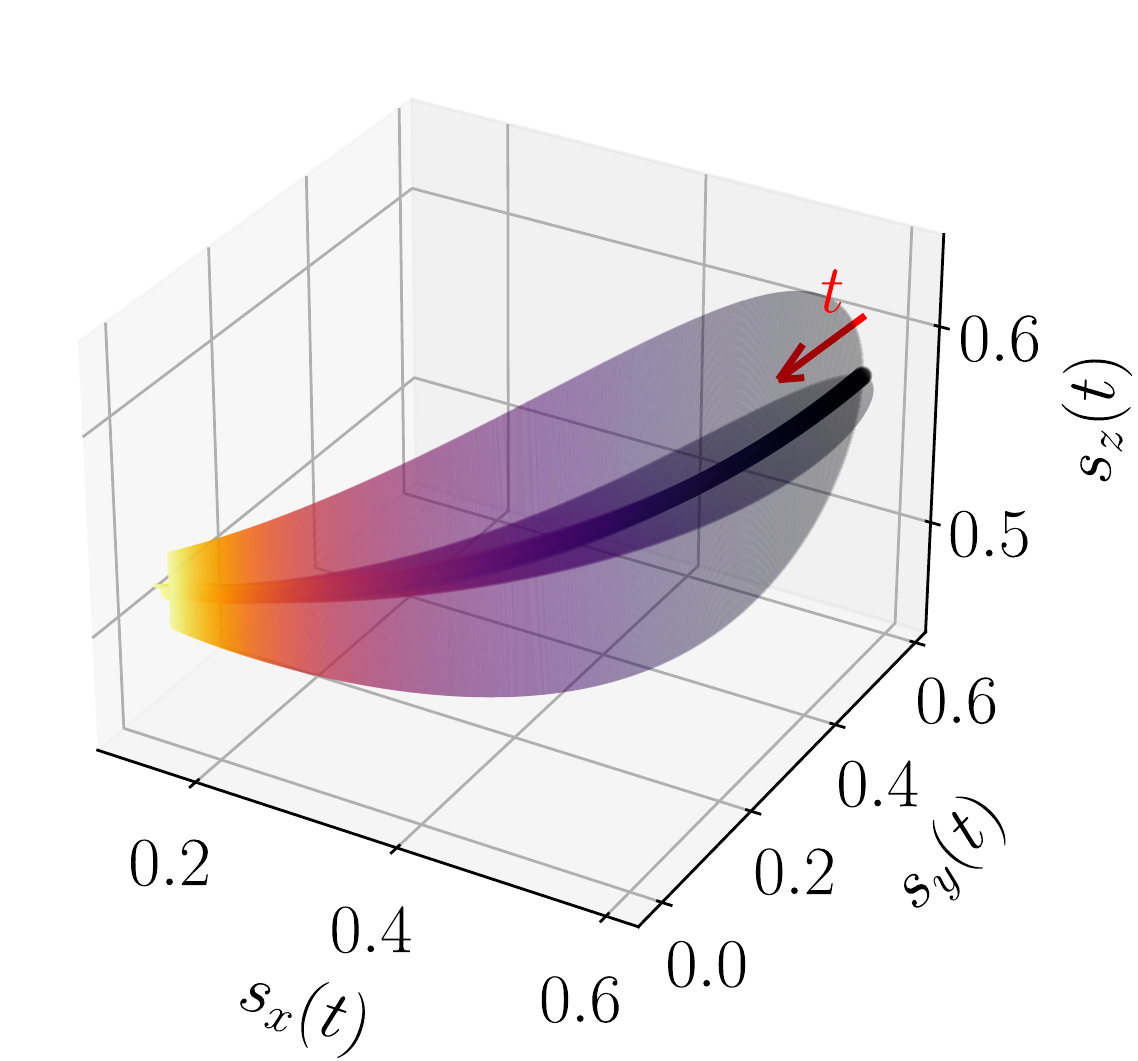}
	\includegraphics[width = .74 \linewidth]{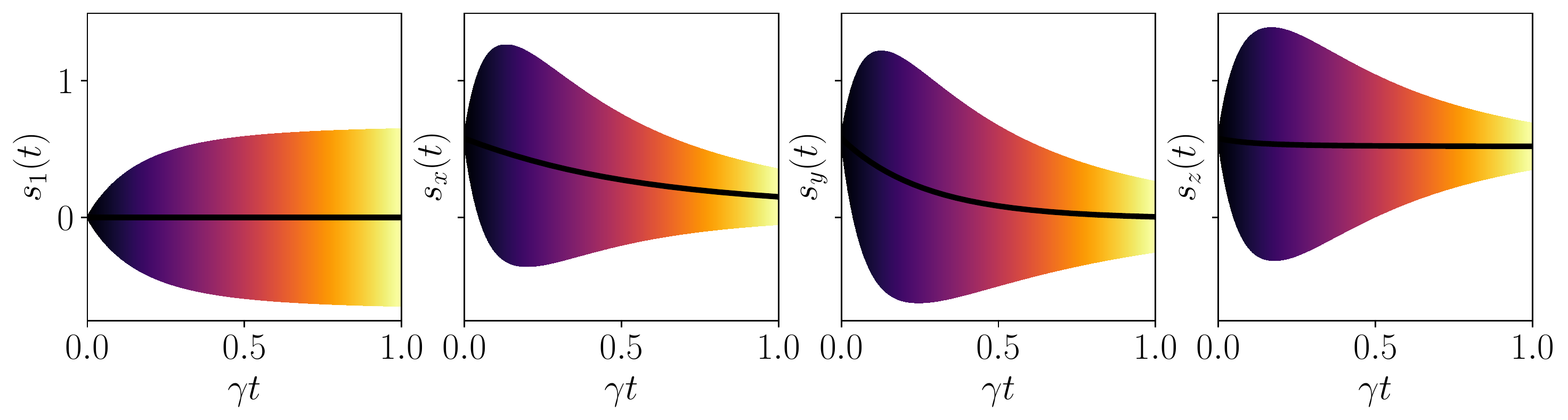}
	\label{fig:evolVarStdDevProj}
	\caption{\textbf{Visualization for the evolution of the stochastic operator variance and its standard deviation for the quantum sLMG model}. (a) Projection over the different spin operators of the noise averaged observable $(\av{\ha A_t}, \ha S_j)$ (solid line) and the SOSD $(\Delta \ha A_t, \ha S_j)$ (error bar), with the flow of time indicated by  the color scale and the red arrow. (b) Projections of the noise-averaged observable (black line) $(\av{\ha A_t}, \ha X)$ and the SOV $(\Delta \ha A_t^2, \ha X)$ (error bar) over $ \ha X \in \{\ha{\mbb 1}, \ha{\be S}\}$. The initial operator is $\ha A=(\ha S_x+\ha S_y + \ha S_z)/\sqrt{3}$. The parameters  are $S=20$ and $\Omega = 1.5$, with $\gamma=0.1$ (upper) or $\gamma = 2$ (lower). The times for which the error bar vanishes correspond to the SOV being orthogonal to the projected operator. \label{fig:visualizeSOV}}
\end{figure*}

We further introduce a notion of Stochastic Operator Standard Deviation (SOSD) as the matrix square root of the SOV, $\Delta \ha A_t=\sqrt{\Delta \ha A_t^2}$. 
The SOV and its deviation are thus illustrated in Fig. \ref{fig:visualizeSOV}. 

\section{Classical stochastic LMG model} \label{sec5}
\subsection{Classical limit of the LMG Hamiltonian}

Let us introduce the SU(2) coherent states as \cite{perelomov_coherent_1986}
\beq
\ket{\zeta} = \frac{e^{\zeta \ha S_+}}{(1+|\zeta|^2)^S}\ket{S, -S},
\eeq
where $\ha S_+=\ha S_x + i \ha S_y$ is the raising operator and $\ket{S, -S}$ is the eigenstate of $\ha S^2$ and $ \ha S_z$ with smallest $z$ component of the spin, namely $\ha S^2 \ket{S, -S}= S(S+1)\ket{S, -S}$ and $\ha S_z \ket{S, -S}= -S \ket{S, -S}$. The coherent states $\ket{\zeta}$ represent a wavepacket with the minimum width allowed by the uncertainty principle localized around $\be n = (\sin \theta \cos \phi, \sin  \theta \sin \phi, \cos \theta)$ where $\zeta = - \tan \frac{\theta}{2}e^{- i \phi}$ and corresponds to a stereographic projection. Following \cite{pilatowsky-cameo_positive_2020}, we define the classical Hamiltonian $H_\ts{lmg}$ as
\beq
H_\ts{lmg} = \lim_{S \rightarrow \infty} \frac{1}{S}\braket{\zeta|\ha H_\ts{lmg}|\zeta},
\eeq
where $S=N/2$. The expectation value of the relevant spin operators between coherent SU(2) states is  \cite{perelomov_coherent_1986}
\beqa
\braket{\be n|\ha S_z|\be n} = - S \cos \theta, \qquad
\braket{\be n|\ha S_x^2|\be n} = S(S-\tfrac{1}{2}) \sin^2 \theta \cos^2 \phi + \tfrac{S}{2}.
\eeqa
The classical LMG Hamiltonian then reads
\beq\label{HlmgThetaPhi}
H_\ts{lmg} = - \Omega \cos \theta - \sin^2 \theta \cos^2 \phi + \mc O(S^{-1}).
\eeq

We then introduce the canonical variables $Q, \; P$ as
\beq
\zeta = \frac{Q - i P}{\sqrt{4 - (Q^2 + P^2)}}= - \tan \frac{\theta}{2} e^{-i \phi},
\eeq
that yield the relations
\begin{align}
	\frac{Q}{\sqrt{4 - Q^2 - P^2}}= - \tan \frac{\theta}{2} \cos \phi, \qquad
	\frac{P}{\sqrt{4 - Q^2 - P^2}}= - \tan \frac{\theta}{2} \sin \phi. 
\end{align}
Inverting theses equations gives 
\beq
\tan \phi = \frac{P}{Q}, \quad \tan^2 \frac{\theta}{2} = \frac{Q^2 + P^2}{4 - (Q^2 + P^2)},
\eeq
that can be substituted in \eqref{HlmgThetaPhi} and yield  
\beq
H_\ts{lmg} = \frac{\Omega}{2}(Q^2 + P^2)-\Omega - \frac{1}{4} (4Q^2-Q^2 P^2 - Q^4).
\eeq
This corresponds to the classical Hamiltonian given in the main text, up to a constant shift in energy.

\subsection{Heuristic reason for stabilization of sLMG}
\begin{figure}[ht]
	\centering
	\includegraphics[width = .5 \linewidth]{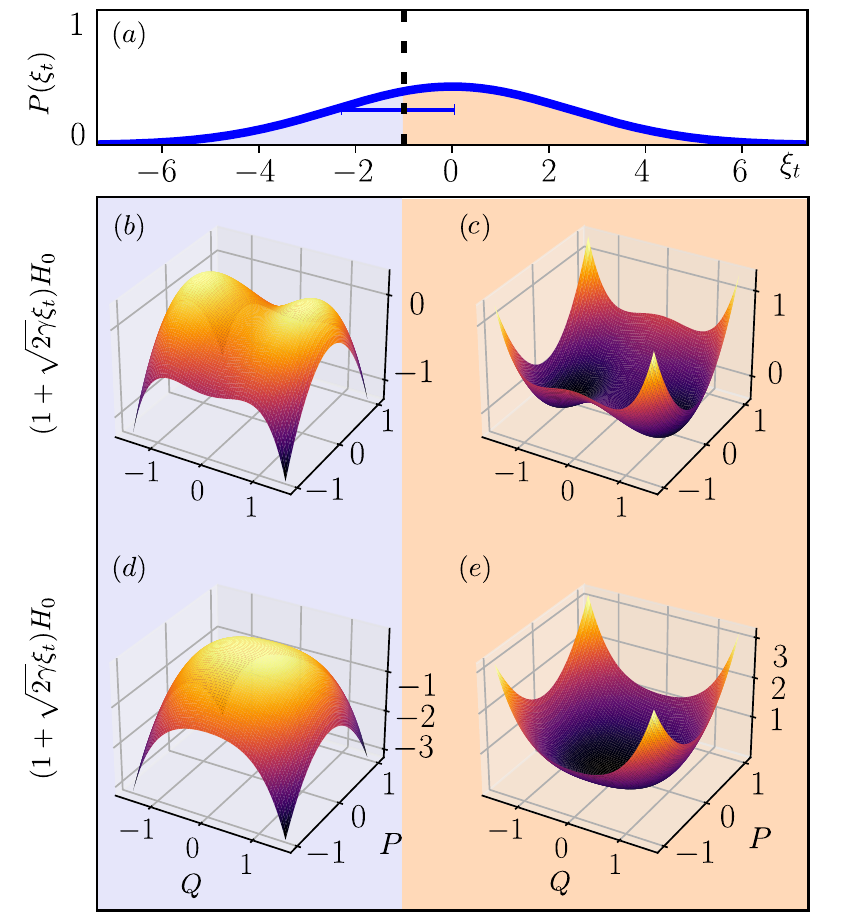}
	\caption{\textbf{Visualization of the stochastic LMG model}. (a) Histogram of the Gaussian white noise. The standard deviation $\sqrt{2\gamma}$ is indicated by the horizontal blue line and the vertical dotted line delimits the sign flip of $(1+\sqrt{2 \gamma}\xi_t)$. LMG Hamiltonian in the double well (b;c) and single well (d;e) phases multiplied by a negative (b;d) and positive (c;e) number corresponding to $(1+\sqrt{2 \gamma} \xi_t)$.}
	\label{fig:sLMG_vis}
\end{figure}

The stabilization we discussed in the main text can be understood in terms of the energy landscape associated with  the stochastic Hamiltonian, illustrated in Fig. \ref{fig:sLMG_vis}. In the limit  that $\sqrt{2 \gamma}\ll 1$, most realizations are as in the noiseless LMG.  The behavior is then close to that of the deterministic model, with the double well (DW) phase showing an unstable point and the single well (SW) phase being stable. In the opposite limit,  $\sqrt{2 \gamma}\gg 1$, almost half of the realizations flip the Hamiltonian since $(1 + \sqrt{2 \gamma}\xi_t)<0$, as shown in Fig. \ref{fig:sLMG_vis} (a). This brings a major difference between the DW and SW phases. In the SW phase, all points in phase space $(Q,P)$ either gain (d)$\rightarrow$(e) or lose (e)$\rightarrow$(d) energy. In the DW phase, however, when going from (b)$\rightarrow$(c), the points inside the wells lose energy while the points outside the wells gain it, and viceversa for (c)$\rightarrow$(b). This difference between the points of phase space provides a rationale for the stochastic stabilization seen for the DW phase---blue region in Fig. \ref{fig:Lyap_sLMG}(b) of the main text.

\subsection{Analytical determination of the Lyapunov exponent for the sLMG model}
We consider a vector variable $\be u_t$ subject to deterministic and fluctuating external perturbations, $\mathbb{A}_d$ and $\mathbb{A}_s(t)$, respectively. Its equation of motion is described by the  stochastic differential equations
\begin{equation}
	\dot{\be{u}}_t = [\mathbb{A}_d + \sqrt{2 \gamma}\mathbb{A}_s(t)] \be u_t.
\end{equation}
Following van Kampen \cite{kampen_stochastic_1992}, we go in the interaction picture  with respect to the deterministic evolution and consider $\be v_t = e^{- \mathbb{A}_d t} \be u_t$. In second-order of $\sqrt{\gamma}$, the average with fixed initial conditions evolves as  
\begin{equation}
	\begin{split}
		&\av{\be v_t} = \be v_0+ 2 \gamma \int_0^t dt_1 \int_0^{t_1} d\tau  
		e^{-t_1 \mathbb{A}_d}\av{\mathbb{A}_s(t_1) e^{\tau \mathbb{A}_d} \mathbb{A}_s(t_1-\tau)}e^{(t_1-\tau)\mathbb{A}_d} \be v_0,
	\end{split}
\end{equation}
valid for $\sqrt{2\gamma}t\ll 1$. We recognize the solution to order $\gamma$ of the linear differential equation, also known as Bourett's integral equation, written back in the original representation as
\begin{equation}
	\label{bourrett}
	{\partial_t} \av{\mathbf{u}_t} = \Big[\mathbb{A}_d  + 2\gamma \int_0^t \av{\mathbb{A}_s(t)e^{\mathbb{A}_d t}\mathbb{A}_s(t-\tau)} e^{-\mathbb{A}_d\tau}d\tau\Big]\av{\mathbf{u}_t}. 
\end{equation} 
This equation is derived assuming the standard rules of calculus, and thus assumes Stratonovich formalism. The latter defines the stochastic integral as ${\int_0^t \delta(t-\tau) f(\tau) d \tau = \frac 1 2 f(t)}$. 
So in the case that $\mathbb{A}_d$ and $\mathbb{A}_s$ commute, and for  $\mathbb{A}_s(t) = \xi_t \mathbb{A}_s$ fluctuating with Gaussian white noise, Eq. \eqref{bourrett} simplifies to 
\begin{align}
	\frac{\partial}{\partial t} \av{\mathbf{u}_t} &= \left(\mathbb{A}_d  - 2\gamma \mathbb{A}_d^2\int_0^t \av{\xi_t \xi_{\tau'}} \notag d\tau'\right)\av{\mathbf{u}_t}
	=\left( \mathbb{A}_d  - \gamma \mathbb{A}_d^2\right)\av{\mathbf{u}_t},
\end{align}
where the change of integration variable $\tau'=t-\tau$ brings the minus sign. In systems exhibiting chaos, the Lyapunov gives the exponential divergence of the trajectory. We interpret this as the maximum eigenvalue of $\mathbb{A}_d  - \gamma \mathbb{A}_d^2$. 

The LMG at the origin, $Q=P=0$, can be linearized into the harmonic oscillator $H=\frac{1}{2}[\Omega P^2 + (\Omega-2)Q^2]$. Hamilton's equation of motion gives $\dot{Q}_t$ and $\dot{P}_t$, from which we can compute the evolution of the quadratic terms as \cite{van_kampen_stochastic_1976}
\begin{equation}
	\frac d {dt} \Bigg( \!\!\begin{array}{c}
		Q_t^2  \\
		P_t^2 \\
		Q_t P_t
	\end{array}
	\!\! \Bigg) {=}\Bigg( \begin{array}{ccc}
		0 & 0 & \Omega  \\
		0 & 0 & -(\Omega{-}2) \\
		- \frac{\Omega-2}{2} & \frac{\Omega}{2} & 0
	\end{array}
	\Bigg)\Bigg( \!\! \begin{array}{c}
		Q_t^2  \\
		P_t^2 \\
		Q_t P_t
	\end{array}
	\!\! \Bigg) \equiv \mathbb{A}_d  \be u_t
	.
\end{equation}
The maximum eigenvalue of this matrix is $\lambda_\ts{lmg} = \sqrt{2 \Omega - \Omega^2}, $ which recovers the Lyapunov exponent at the origin in the noiseless LMG model \cite{pilatowsky-cameo_positive_2020}.
For the sLMG, we add noise in the energy scale and consider an evolution dictated by 
\beq
\dot{\be u}_t = \mathbb{A}_d (1 + \sqrt{2 \gamma} \xi_t) \be u_t, 
\eeq
where $\xi_t$ is Gaussian white noise. 
The  maximum eigenvalue of $\mathbb{A}_d  - \gamma \mathbb{A}_d^2$  thus gives the average Lyapunov exponent as  
\beq
\lambda = \sqrt{2 \Omega - \Omega^2} - \gamma (2 \Omega - \Omega^2).
\eeq

\section{Details on the numerical solutions} \label{sec6}
\subsection{Vectorization}

The formal solution to the master equation reads $\av{\ha A_t} = e^{\mc L\dg t}[\ha A]$, where $\mc L\dg [\bullet]$ is the adjoint Liouvillian superoperator. To numerically solve it, we apply vectorization by which operators $\ha A_t$ become vectors in a larger Hilbert space $|\ha A_t)$. Superoperators become operators over this larger Hilbert space through the mapping 
\begin{equation}
	\ha X \ha A_t \ha Y \rightarrow (\ha X \otimes \ha Y^T)|\ha A_t),
\end{equation}
therefore the vectorized Lindbladian reads
\begin{equation}
	\mc L\dg[\bullet] \rightarrow i \ha H_0 \otimes \mbb 1 - i \mbb 1 \otimes \ha H_0^T + \gamma (2 \ha L \otimes \ha L^T - \ha L^2 \otimes \mbb 1 - \mbb 1 \otimes (\ha L^2)^T).
\end{equation}

\subsection{Solver of Stochastic Differential Equations}
We solve the SDE using the \textit{explicit order 1.0 strong scheme} \cite{kloeden_numerical_1992}, which briefly consists of the following. Consider a It\=o SDE of the form
\beq
\mr d X_t = a(X_t)\mr dt + b(X_t) \mr d W_t,
\eeq
where $X_0$ is the initial condition. Let $Y_n$ be the solution at time $t_n = n \delta,$ where $\delta$ is the time-step. We first set the initial condition $Y_0 = X_0$ and compute recursively the solution as
\begin{align}
	Y_{n+1}=&Y_n + a_n \delta + b_n \Delta W_n 
	+\frac{1}{2 \sqrt{\delta}}(b(\Upsilon_n)-b_n)((\Delta W_n)^2-\delta) , 
\end{align}
where $\Upsilon_n = Y_n + a_n \delta + b_n \sqrt{\delta}$, $a_n = a(Y_n), \; b_n = b(Y_n)$, and $\Delta W_n= W_{n+1}-W_n$ are independent identically distributed normal random variables with zero average and variance $\delta$.

\end{document}